\documentclass[twocolumn,showpacs,preprintnumbers,amsmath,amssymb]{revtex4-1}

\usepackage{graphicx}
\usepackage{dcolumn}
\usepackage{bm}
\usepackage[dvipdfmx]{hyperref} 

\begin{document}


\title{Cooling dynamics of a single trapped ion via elastic collisions with small-mass atoms} 

\author{Shinsuke Haze$^{1}$}
\email{haze@ils.uec.ac.jp}
\author{Mizuki Sasakawa$^{1}$}
\author{Ryoichi Saito$^{1}$}
\author{Ryosuke Nakai$^{1}$}
\author{Takashi Mukaiyama$^{2,3}$}%

\affiliation{
$^{1}$\mbox{Institute for Laser Science, University of Electro-Communications, Chofugaoka, Chofu, Tokyo 182-8585, Japan}\\
$^{2}$\mbox{Graduate School of Engineering Science, Osaka University, 1-3 Machikaneyama, Toyonaka, Osaka 560-8531, Japan}\\
$^{3}$\mbox{PRESTO, Japan Science and Technology Agency, Honcho, Kawaguchi, Saitama 332-0012, Japan}\\
}

%
%
%


\begin{abstract}
We demonstrated sympathetic cooling of a single ion in a buffer gas of ultracold atoms with small mass.
Efficient collisional cooling was realized by suppressing collision-induced heating.
We attempt to explain the experimental results with a simple rate equation model and provide a quantitative discussion of the cooling efficiency per collision.
The knowledge we obtained in this work is an important ingredient for advancing the technique of sympathetic cooling of ions with neutral atoms.
\end{abstract}
\maketitle


Sympathetic cooling, where we thermally contact two distinct systems at different temperatures, is an effective method for cooling an object to a desired energy regime.
Nowadays, this is commonly used in the field of low-temperature physics for producing a cold sample for a molecular beam or degenerate atomic gases.
The elemental mechanism of this technique extracts energy from a target object through interaction (normally by collisions) with a coolant system.
In cooling of translational motion, for example, an exchange of moment between the two systems can remove kinetic energy from the thermal system, and eventually they reach thermal equilibrium, resulting in cooling of the target object.

To introduce an ultracold atomic gas as a coolant for trapped ions is attractive, since collisions with ultracold atoms enable efficient cooling of a number of vibrational modes simultaneously.
It is beneficial for many applications, for example, continuous cooling of ion qubits in quantum information processing.
In addition, this method has been proven to be effective for the cooling of atomic or molecular systems, especially when no conventional laser cooling transition is accessible, as demonstrated in rotational-vibrational cooling of molecular ions \cite{Rellergert,Hansen}.

In buffer-gas cooling of a charged particle, however, the situation is not as simple as in the usual schemes, e.g., evaporation or sympathetic cooling in a mixture of neutral gases.
This is simply attributed to the dynamics of trapping it in a radiofrequency (RF) trap, where slow (secular) and rapid (micro) motion are superimposed on the motion of the ion.
The main point is that an abrupt interception of coherent ion motion by an atom-ion collision complicates the kinetics of the ion.
Importantly, an ion can be either cooled or even heated, depending on the instantaneous phase of micromotion at the moment of a collision, which prevents efficient collisional cooling \cite{Cetina}.
In addition, this peculiar feature modifies the energy distribution of an ion by inducing a deviation from a normal (Maxwell-Boltzmann) distribution to a super-statistical (so-called Tsallis) distribution accompanied with a power-law tail in a high-energy region \cite{DeVoe, Zipkes, Rouse}.
These subjects have been pointed out since the early stages of ion trapping experiments \cite{Major}, and they were recently revisited again in conjunction with the rapidly growing field of ultracold atom-ion hybrid systems \cite{Tomza}.

A key parameter for characterizing the kinetics of the ion in a buffer gas is the atom-to-ion mass ratio $m_{\rm{a}}$/$m_{\rm{i}}$.
Here, $m_{\rm{a}}$ and $m_{\rm{i}}$ denote mass of an atom and an ion, respectively.
Recent study has revealed that the heating effect by collisions with atoms strongly depends on the mass ratio and a critical value that determines the boundary for cooling and heating, which is suggested to be $m_{\rm{a}}/m_{\rm{i}}\sim$1.0 \cite{Zipkes, Chen, Holtkemeier}.
As a consequence, small-mass atoms are found to be advantageous for suppressing such heating effects.

\begin{figure}[t]
\includegraphics[width=8.5cm]{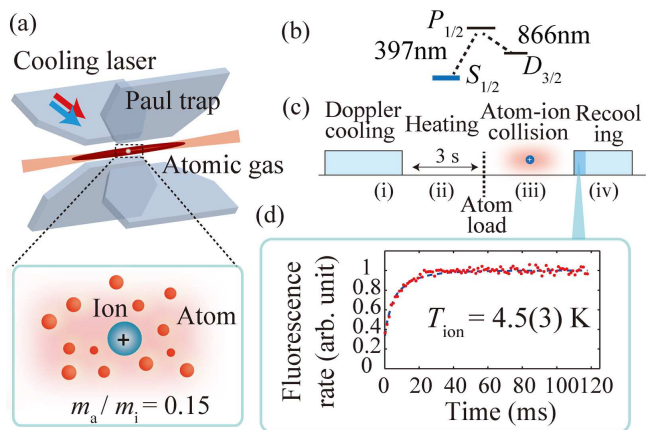}\\
\caption{\label{setup} (Color online) (a) Sketch of atom-ion hybrid apparatus for sympathetic cooling.
A single ion is captured in a linear RF trap and immersed in an ultracold atomic gas of Li atoms.
As the ion undergoes elastic collisions with atoms, it loses kinetic energy.
(b) Relevant energy levels of Ca$^{+}$ ion.
The $S_{1/2}$-$P_{1/2}$ transition is used for both Doppler cooling and
ion thermometry.
(c) Experimental sequence.
The atom-ion sympathetic cooling sequence includes (i)~Doppler cooling for initialization, (ii)~energy tuning of the ion, (iii)~sympathetic cooling via collisions, and (iv)~temperature probing.
(d) Ion fluorescence during recooling.
The plot is an example of a transient signal during Doppler recooling plus a fitting curve (dashed line).
The temperature is 4.5(3)~K. } 
\end{figure}

In this work, we experimentally studied the kinetics of a single ion in a buffer gas of ultracold atoms.
We introduced an atomic gas of Li atoms ($m_{\rm{Li}}$=6) as a coolant for a Ca$^{+}$ ion ($m_{\rm{Ca^{+}}}$=40) in an RF trap.
See Fig.~\ref{setup}(a).
Li has the smallest mass among the atomic species that are amenable to conventional laser cooling.
The mass ratio here is $m_{\rm{a}}$/$m_{\rm{i}}$=0.15, and this is sufficiently smaller than the critical value.
The energy threshold for realizing quantum scattering in atom-ion systems is also strongly affected by the mass as given by $E_{\rm{th}}$=$\frac{\hbar^{4}}{2C_{4}\mu^{2}}$,
where $\mu$ is an atom-ion reduced mass. 
Therefore, the mixture composed of Li atoms presented here is considered suitable for ultracold atom-ion physics, as well as the recently realized Li-Yb$^{+}$ mixture \cite{Joger}.

In the context of the kinetics of an ion in a buffer gas, essential theoretical work has been reported \cite{DeVoe, Cetina, Zipkes, Chen, Holtkemeier, Rouse}.
Regarding experimental work, cooling behavior in an atom-ion hybrid system has been presented in several pioneering works \cite{Goodman, Zipkes2, Harter, Sivarajah}, and evidence of collisional stability \cite{Ravi}, blue-sky bifurcation \cite{Schowalter}, cooling effect in heavy-atom and light-ion systems \cite{Dutta} and cooling by resonant charge exchange \cite{Dutta2} has been demonstrated.
In addition, power-law statistics and the role of the atom-ion mass ratio was experimentally revealed in an equal-mass system \cite{Ziv}.
However, an experimental approach to verify buffer-gas cooling with an ultracold gas with small mass has not been performed until now. 

Here, we address the temperature evolution of a Ca$^{+}$ ion undergoing collisions with Li atoms.
As far as we know, this is the first demonstration of sympathetic cooling with Li atoms.

We trapped a single Ca$^{+}$ ion in a linear RF trap [Fig.~\ref{setup}(a)], where the ion was confined in an electric potential generated by static and oscillating RF fields of 25.4~MHz.
The trapping frequency was ($\omega_{r}$, $\omega_{a}$)=2$\pi\times$(1,500, 810)~kHz,
where $\omega_{r}$ and $\omega_{a}$ represent the radial and axial trapping frequencies, respectively.
The Mathieu parameter $q$ is 0.17, which also affects the dynamics of the ion in a buffer gas, as well as the mass ratio; this value is within a stable cooling realm \cite{Chen}.
The ion was Doppler-cooled to 1.5~mK using the $S_{1/2}$-$P_{1/2}$ transition [Fig.~\ref{setup}(b)].

Li atoms were trapped in an optical dipole trap as shown in Fig.~\ref{setup}(a).
The radial and axial trap frequencies were ($\omega_{r}$, $\omega_{a}$)=2$\pi\times$(1,010, 4.5)~Hz, respectively.
The typical number of atoms was 20$\times$10$^{3}$, and the temperature was 4.5~$\mu$K. 
Atoms were spatially overlapped with the ion for atom-ion thermalization.
A detailed explanation of our apparatus can be found elsewhere \cite{Haze2, Saito}.

The experimental sequence involved the following steps, as shown in Fig.~\ref{setup}(c) .
First, an ion was (i)~Doppler cooled, and then (ii)~the temperature was tuned (heated) to 6--7~K by turning off the cooling laser for 3~s.
Subsequently, (iii)~an atomic cloud was loaded to induce atom-ion collisions, and finally (iv)~the cooling laser was promptly turned on again for recooling.
The atoms were released at the end of step~(iii).
We recorded the photon emissions from the ion in step~(iv) to obtain the transient dynamics of the fluorescence of the ion after turning on the cooling laser.
We were able to estimate the temperature of the ion by analyzing the signal during this Doppler recooling (DRC) \cite{Epstein, Ziv}.
We repeated this sequence 100--500 times in determining the temperature.
Since this thermometry method has a good sensitivity in a temperature range of hundreds of mK to 10~K \cite{Sikorsky}, we intentionally heated the ion in step (ii).

\begin{figure}[t]
\includegraphics[width=8.5cm]{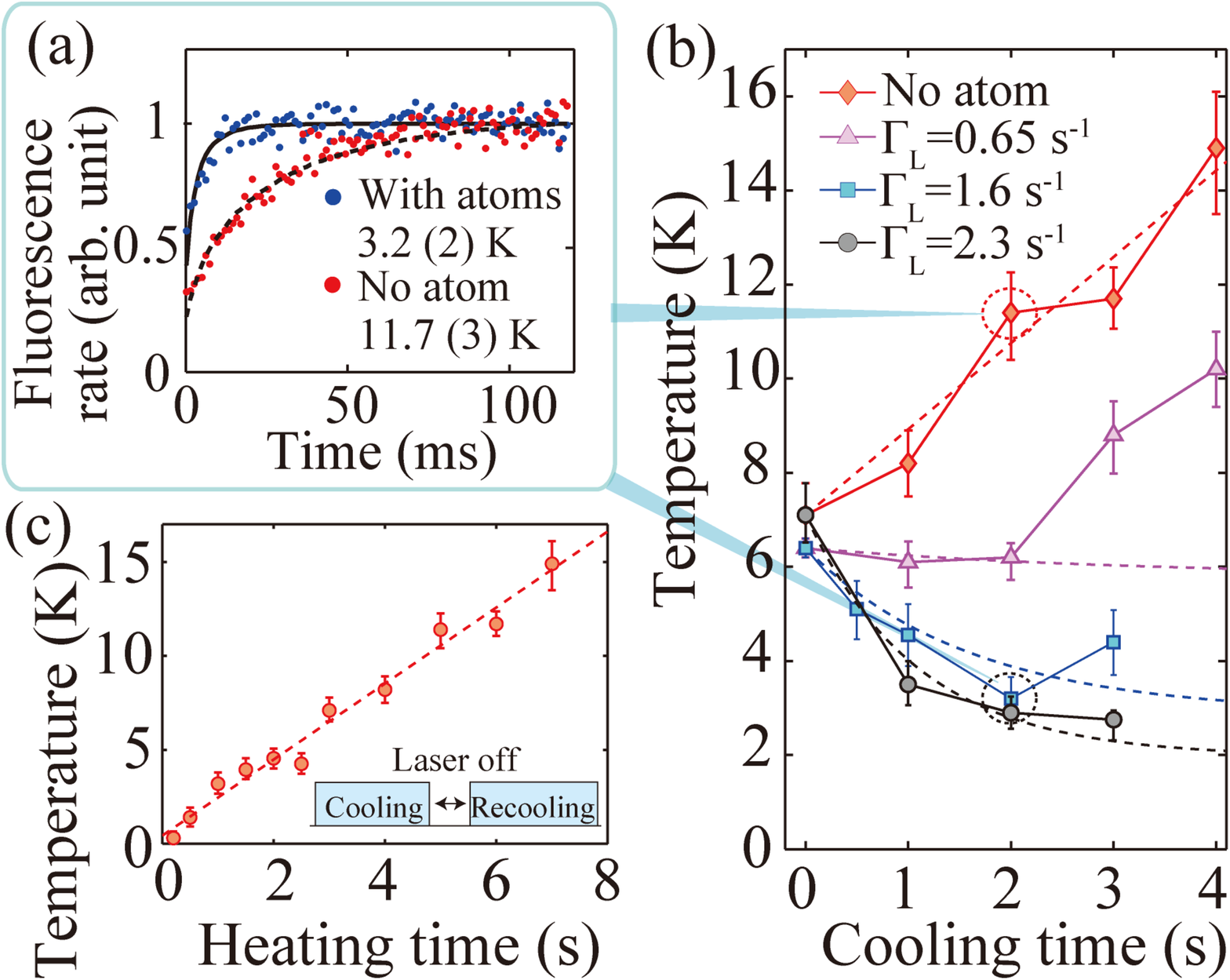}\\
\caption{\label{dynamics} (Color online) 
(a) Two examples of DRC signals with and without buffer-gas cooling.
(b) Cooling dynamics of Ca$^{+}$ ion in Li gas.
The temperature of the ion at various collisional cooling times is shown.
We plot data for four different collision rates: $\Gamma_{\rm{L}}$ = 2.3~s$^{-1}$ (black circle), 1.6~s$^{-1}$ (blue square), 0.65~s$^{-1}$ (magenta square), and no atom (red diamond).
The dashed curve for each rate is a solution of the analytical model discussed in the text.
(c) Intrinsic heating dynamics measured in this setup.
To perform this measurement, we probed the temperature of the ion while changing the heating time.
The sequence is shown in the inset.
The dashed line is a fit with a linear line.
}
\end{figure}

In Fig.~\ref{dynamics}(a), two examples of DRC signals in the presence and absence of buffer-gas cooling are shown.
During DRC, the time to reach a steady fluorescence rate depends on the initial temperature (i.e., a hotter ion takes a longer time).
We observed a quick recovery of the fluorescence rate for a sympathetically cooled ion; in contrast, a delay was seen for non-cooled ions, evidencing the cooling effect by the buffer-gas atoms.
The temperature was 3.2~K and 11.7~K with and without buffer gas, respectively.
To derive these values, we fitted the fluorescence data with a calculated curve based on \cite{Wesenberg}.
In this calculation, the temperature of the ion is a fitting parameter, while the other parameters, such as laser detuning and saturation factor, are fixed. 
Here, a thermal distribution of ion energy is presumed instead of a power-law distribution.
This is considered natural, since the mass ratio is small and the ion kinetics are expected to be more close to a normal distribution.

We performed the above measurements while changing the cooling period in step~(iii) for further investigation of this cooling behavior. 
The result is shown in Fig.~\ref{dynamics}(b), where the evolution of the temperature of the ion in a buffer gas is displayed.
Each of the four plots corresponds to a result with a collision rate of $\Gamma_{\rm{L}}$=2.3~s$^{-1}$, 1.6~s$^{-1}$, 0.65~s$^{-1}$, and 0~s$^{-1}$ (no atom).
Starting from 6.4~K, the temperature of the ion was dynamically changed as collisions proceeded, and different cooling time constants were observed at different collision rates.
To vary the collision rate, we changed the overlap between the ion and atomic cloud, which is basically the atomic density at the position of the ion, in a controlled manner by shifting the position of the dipole trap using a motorized positioner.

During buffer-gas cooling, the ion was constantly in the $S_{1/2}$ state, since the cooling laser was turned off.
Therefore, the observed temperature change is purely attributed to collisions with atoms.
We rarely observed an ion loss during this measurement; basically, inelastic collision, such as charge-exchange or molecular formation, was negligible.
This robustness against particle loss is most favorable for buffer-gas cooling.
Meanwhile, once the ion was excited to an upper state, i.e., $P_{1/2}$ or the metastable $D$ state, the ion was susceptible to a reactive collision.
This reactivity, in contrast to the $S_{1/2}$-state ion, is practically useful for determining the collision rate.
In this work, we calibrated the collision rate utilizing the loss mechanism of an ion due to the charge-exchange process (Ca$^{+}$ + Li $\rightarrow$ Ca + Li$^{+}$) in the excited state \cite{Saito}.
The collision rate $\Gamma_{\rm{L}}$ in this work is equivalent to the rate for Langevin collision, which is derived from an independent measurement of charge-exchange loss of ions in the presence of a cooling laser.

To model the cooling dynamics, we introduce a simple rate equation written as,
\begin{eqnarray}
\label{eq:model}
\frac{ \mathrm{d}} {\mathrm{d}t}T_{\rm{ion}}=-\eta T_{\rm{ion}} + R_{\rm{heat}}.
\end{eqnarray}
The first term corresponds to collisional cooling with $\eta$ being the cooling efficiency per unit time.
Here, an ion is assumed to hit an atom at rest, because the atomic temperature is negligibly small compared to that of the ion.
The second term is heating of an ion, which leads to a linear increase of the temperature over time.
$R_{\rm{heat}}$ is the heating rate.
Each of the dashed curves in Fig.~\ref{dynamics}(b) represents the solution of Eq.~(\ref{eq:model}).
The analytical form of this curve is given from Eq.~(\ref{eq:model}) as
\begin{eqnarray}
\label{eq:T_ion}
T_{\rm{ion}}\left( t \right)=\left( T_{0}-\frac{R_{\rm{heat}}}{\eta}\right) e^{-\eta t}+\frac{R_{\rm{heat}}}{\eta},
\end{eqnarray}
where $T_{0}$ is the initial temperature at $t$=0.
We performed fitting to the data to draw the curves.
In fitting, $\eta$ was used as a fitting parameter and $R_{\rm{heat}}$ as a fixed value.
$R_{\rm{heat}}$ is obtained from the experimental result in Fig.~\ref{dynamics}(c), which indicates the heating dynamics of a single ion.
The heating rate of 1.9~K/s is derived from a linear fit (dashed line). 
In this experiment, we measured the temperature after a variable heating time with no cooling laser and no atoms.
The experimental sequence is illustrated in Fig.~\ref{dynamics}(c), where we shut off the cooling laser for a certain time and turn the cooling laser back on to recool for thermometry. 
The measured heating rate here is relatively higher than the empirically derived values \cite{Brownnutt}. 
We estimate this is primary attributed to anomalous heating due to electrical noise caused by contamination of the electrode surface.

\begin{figure}[t]
\includegraphics[width=8cm]{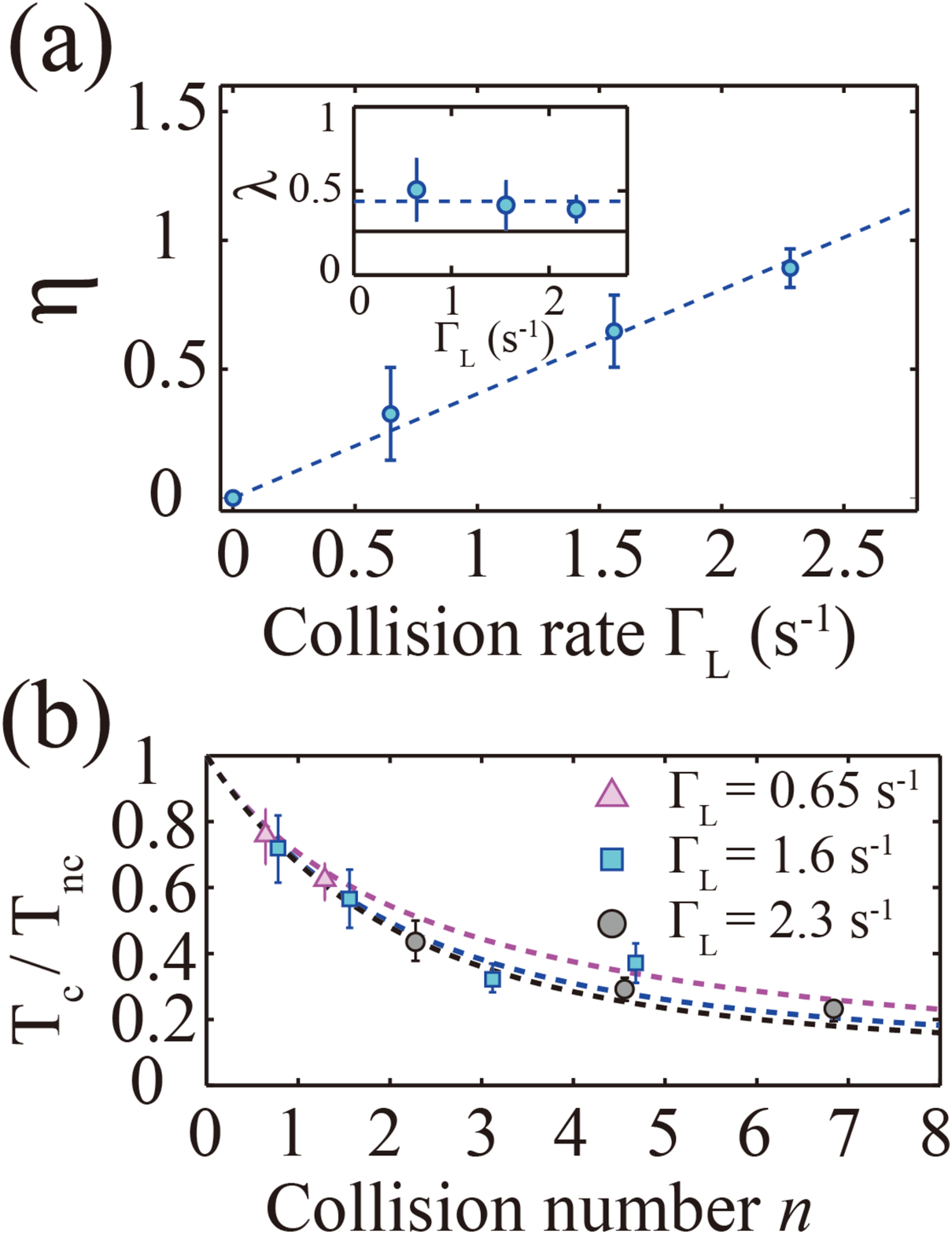}\\
\caption{\label{analysis} (Color online) (a) Cooling efficiency $\eta$ at different collision rates. 
The dashed line is a result of fitting with a linear function.
The inset plots $\lambda$, which is found to be constant at various collision rates.
The dashed line is the average of the data.
The solid line is the value predicted for $\lambda$ based on the assumption of head-on collisions ($\lambda$=0.26).
(b) Energy reduction $T_{\rm{cool}}/T_{\rm{heat}}$ vs. collision number $n$.
The dashed line is the solution of Eq.~(\ref{eq:model}) with $\lambda$=0.42.
 } 
\end{figure}

In Fig.~\ref{analysis}(a), we plot $\eta$, which gives the best fitted curve for each collision rate.
The dashed line is a result of fitting with a linear function.
Here, we define the cooling efficiency per single collision as $\lambda=\eta/\Gamma_{\rm{L}}$.
$\lambda$ for each collision rate (except for case with no atom) is plotted in the inset of Fig.~\ref{analysis}(a).
We found that $\lambda$ has constant values for each of the three different collision rates and the average value of $\lambda$ (dashed line) is 0.42(12).
We estimate that the two data points with $\Gamma_{\rm{L}}$ = 0.65~s$^{-1}$ [Fig.~\ref{dynamics}(b)] anomalously deviated from the prediction mainly due to inefficient cooling in the high-temperature region.
Since the slope of the temperature increase from 2 to 4~s matches the heating rate in the case of no cooling (red dashed line), the heating mechanism is expected to be dominant here.
One possibility for this behavior is a reduction of cooling efficiency due to a loose overlap with atoms, which originates from the large orbit of the ion at high temperature.
Based on this reasoning, we excluded these two data points in this analysis. 

To trace this cooling dynamic with respect to the number of collisions, we reanalyzed the data in Fig.~\ref{dynamics}(b).
For this purpose, we plot the energy removal defined by $T_{\rm{c}}/T_{\rm{nc}}$ in terms of the collision number $n$ in Fig.~\ref{analysis}(b).
Here, $T_{\rm{c}}$ and $T_{\rm{nc}}$ represents the temperature of the ion in Fig.~\ref{dynamics}(b) after a certain period with cooling and no cooling, respectively.
In Fig.~\ref{analysis}(b), we also plot the fitting curve calculated from Eq.~(\ref{eq:model}) for each collision rate.
Once $\lambda$ is obtained from the fitting in Fig.~\ref{dynamics}(b) and Fig.~\ref{analysis}(b), the cooling capability per collision can be estimated.
The motional energy of the ion in the absence of heating is reduced as a factor of $e^{-\lambda}$ for one collision, which is derived from Eq.~(\ref{eq:T_ion}) for $R_{\rm{heat}}$=0.
Thus, the expected energy extraction factor is now calculated to be 0.65 (=$e^{-0.42}$). which means 35\% of the motional energy can be removed with one collision.

Here, we attempt a quantitative evaluation of the obtained cooling efficiency.
The largest contribution for the cooling is from Langevin collisions.
In the Langevin model, a direct nuclear-nuclear collision occurs at a short distance; therefore, we simply regard the collision as a binary head-on collision between an atom and an ion.
Then, the kinetic energy of the ion after one elastic collision can be calculated as $\frac{m_{\rm{i}}^{2}+m_{\rm{a}}^{2}}{(m_{\rm{i}}+m_{\rm{a}})^{2}}$, which is derived as a consequence of averaging over the scattering angle after a collision.
In the case of Li-Ca$^{+}$ collisions, the motional energy is estimated to be reduced by a factor of 0.77 with a single collision, and the corresponding cooling efficiency is calculated as $\lambda$=0.26, as plotted by the solid line in the inset in Fig.~\ref{analysis}(a).
However, the experimental result for $\lambda$ is larger than the value obtained from this model.
Although the uncertainty in determining cooling efficiency (1 standard deviation) is in a comparable regime with this value, this result seems to suggest an additional contribution from other cooling mechanism.
One possibility is atom-ion scattering with a large impact parameter (glancing collision).
On the other hand, the resulting cooling efficiency $\eta$ is linearly dependent on $\Gamma_{\rm{L}}$, as we see in Fig.~\ref{analysis}(a).
This suggests that the cooling effect due to glancing collisions can also be linearly scaled with respect to $\Gamma_{\rm{L}}$.
This finding is consistent with the result of the numerical examination in \cite{Chen}, where the scattering cross-section of glancing and Langevin collisions were argued as approximately comparable, that is, $\sigma_{\rm{g}}\approx\sigma_{\rm{L}}$.
An optimal mass ratio for sympathetic cooling is determined by the interplay between collisional cooling and micromotion effect.
For example, we are able to estimate an optimal value by adopting an analytical method developed by Chen et al. \cite{Chen}. 
Following their model, we find that the cooling efficiency is expected to be maximized at $m_{\rm{a}}/m_{\rm{i}}$=0.38, and the cooling turns to heating at $m_{\rm{a}}/m_{\rm{i}}$=1.25 at our current experimental condition in RF trap (q=0.17). 
On the other hand, the final temperature in atom-ion hybrid trap is also fundamentally limited by the mass ratio as revealed by Cetina et al. \cite{Cetina}, where the residual heating is mitigated by introducing large mass imbalance. 
Therefore, a small-mass atom such as Li is ultimately favorable for reaching an ultracold regime in an atom-ion hybrid system.

In summary, we experimentally demonstrated sympathetic cooling of a single trapped ion using an ultracold buffer gas of Li atoms.
Since the cooling efficiency can be consistently explained with a simple model, the collision-induced heating, which is an issue in atom-ion sympathetic cooling, is not significant, as manifested in a small-mass and heavy-ion system at this temperature scale.
As we expected, the major part of the collisional cooling was provided by the Langevin process, which brings a large momentum transfer, but glancing scattering was also involved in the cooling mechanism in this system.
Currently, our measurement is restricted to a temperature of a few kelvin, primarily 
because the thermometry method has sensitivity in a limited energy regime (typically from hundreds of mK up to tens of K \cite{Sikorsky}).
For example, a different scheme using a coherent excitation with a narrow line transition \cite{Ziv}, will enable entering a lower energy scale.
Our next objective is to affirm the cooling efficiency in this system by a numerical calculation using, for example, molecular dynamics or Monte-Carlo simulations.

This work was supported by the Japan Society for the Promotion of Science, Grants-in-Aid for Scientific Research (KAKENHI, Grant Nos. 26287090, 24105006, 15J10722, and JP16J00890), and the Precursory Research for Embryonic Science and Technology (PRESTO) program of the Japan Science and Technology Agency.

\end{document}